\documentclass[
               twocolumn,
               noshowpacs,          
               nopreprintnumbers,     
               aps,                 
               prd,                 
               letter,             
	       superscriptaddress,  
               nofootinbib,         
               tightenlines,        
               floats,floatfix    
               ]{revtex4}

\usepackage{amsmath, amsfonts, amsthm, amssymb, graphicx}
\usepackage{color}
\usepackage[utf8]{inputenc}
\usepackage{lipsum}
\usepackage{centernot}
\def\be{\begin{equation}}
\def\ee{\end{equation}}
\def\ba{\begin{eqnarray}}
\def\ea{\end{eqnarray}}
\def\beastar{\begin{eqnarray*}}
\def\eeastar{\end{eqnarray*}}       

\def\bdm{\begin{displaymath}}
\def\edm{\end{displaymath}}

\def\bq{\begin{quote}}
\def\eq{\end{quote}}

 at 10truept

\newcommand{\beq}{\begin{equation}}
\newcommand{\eeq}{\end{equation}}
\newcommand{\bea}{\begin{eqnarray}}
\newcommand{\eea}{\end{eqnarray}}
\newcommand{\beqa}{\begin{eqnarray}}
\newcommand{\eeqa}{\end{eqnarray}}

\def\V{{\cal V}}
\def\I{{\cal I}}

\def\ltap{\ \raise.3ex\hbox{$<$\kern-.75em\lower1ex\hbox{$\sim$}}\ }
\def\gtap{\ \raise.3ex\hbox{$>$\kern-.75em\lower1ex\hbox{$\sim$}}\ }
\def\gl{\ \raise.5ex\hbox{$>$}\kern-.8em\lower.5ex\hbox{$<$}\ }
\def\roughly#1{\raise.3ex\hbox{$#1$\kern-.75em\lower1ex\hbox{$\sim$}}}

\begin{document}

\title{Sequestering effects on and of vacuum decay}

\author{Nemanja Kaloper} 
\email{kaloper@physics.ucdavis.edu}
\affiliation{Department of Physics, University of California, Davis, CA95616, USA} 
\author{Antonio Padilla} 
\email{antonio.padilla@nottingham.ac.uk}
\affiliation{School of Physics and Astronomy, 
University of Nottingham, Nottingham NG7 2RD, UK} 
\author{David Stefanyszyn} 
\email{ppxds1@nottingham.ac.uk}
\affiliation{School of Physics and Astronomy, 
University of Nottingham, Nottingham NG7 2RD, UK}

\date{\today}

\begin{abstract}
We consider phase transitions and their contributions to vacuum energy in the manifestly local theory of vacuum energy sequestering. We demonstrate that the absence of instabilities imposes constraints on the couplings of gravitating and non-gravitating sectors, which can be satisfied in a large class of models. We further show by explicit construction that the vacuum energy contributions to the effective cosmological constant in the descendant vacua are generically strongly suppressed by the ratios of spacetime volumes of parent and descendant geometries. This means that the cosmological constant in de Sitter descendant vacua remains insensitive to phase transitions which may have occurred in the course of its cosmic history.

\end{abstract}
\maketitle

\section{Introduction}

`The cosmological constant problem' follows from the universality of gravity and the quantum generation of vacuum energy by virtual particles
(see, e.g. \cite{zeldovich,wilczek,wein,pol,cliff,me}). Even the geometry of space-time in vacuum must be curved, with the curvature set by vacuum energy density. Cosmological observations constrain it to be about 10 billion lightyears, implying that the scale of the vacuum energy density is about milli-electron volt. However theoretical estimates exceed this value manyfold, giving energy density of the vacuum possibly as high as Planckian scales, some 120 orders of magnitude too large.

This huge vacuum energy could be cancelled in a world which is either supersymmetric and/or conformally invariant. In those cases, matter spectra `conspire' to exactly cancel vacuum energy as dictated by symmetry. But our world is neither supersymmetric nor conformal at scales below TeV. An alternative has been to look for a dynamical adjustment of vacuum energy, where a degree of freedom `soaks' it all up and somehow prevents it from gravitating. A problem with realizing this in quantum field theory (QFT) is summarized by Weinberg's no-go theorem \cite{wein} which prohibits such adjustment in any standard QFT coupled to gravity. 

Such a desperate state of affairs prompted work on modifying gravity itself. Typically changing gravity yields fast instabilities and nonlocal/acausal behavior, which conflict the observations. However, recently we have proposed a maximally minimal modification of standard GR involving global gauge variables such as the total spacetime volume \cite{KP1,KP2,KP3,KPSZ}, which cancels all matter vacuum energy contributions from gravitational equations. The setup uses some of the ideas advocated earlier by \cite{andrei,tseytlin}. In this proposal, {\it all} quantum-generated vacuum energy contributions from a protected matter sector cancel completely from gravitational equations of motion. The only vacuum energy which sources gravity is a renormalized vacuum energy, which in our proposal is automatically radiatively stable. 

The numerical value of this quantity is {\it not} determined by the theory. However, this is fully consistent with the spirit of renormalization in QFT. Indeed, in QFT any UV sensitive physical quantities in the theory (particle masses, couplings) must be renormalized: an infinity is subtracted by a bare counterterm, and a boundary condition for the finite remainder picked at some scale $\mu$. Since the scale $\mu$ is completely arbitrary, after the subtraction one is left with a family of theories characterized by the arbitrary subtraction point $\mu$. 
The only way to determine the numerical value of this quantity is to match it to an observation. Once this is done, one can go on and make predictions about all the other physical quantities which are not UV sensitive. 

The physical cosmological constant is a UV sensitive physical variable, including quartically divergent loop-generated contributions from QFT, which are cancelled by the bare counterterm: Einstein's original `classical' $\Lambda$. The net result is finite {\it by design} because it is what sources the background vacuum energy, but its value must be measured. The trick is, 
how does one measure the cosmological constant? The very notion implies the statement that this quantity is known with arbitrary precision - by being a {\it constant}. But because the physical cosmological constant is a space-time filling quantity, this means that it must be measured over all spacetime (this has been noted also in \cite{degrav}). This deep subtlety is normally completely ignored in classical GR since this measurement, while being nonlocal, is ``preordained" by the postulate that the cosmological constant of a particular value is given in the initial data for Einstein's equations. In QFT this just doesn't work.

In our case \cite{KPSZ}, we can write the effective Einstein's equations, $G^\mu{}_\nu = \tau^\mu{}_\nu - \Lambda_{eff} \delta^\mu{}_\nu$, where $\tau^\mu{}_\nu$ 
is the matter stress energy of localized sources (ie., with vacuum energy subtracted off). What must be measured is the residual cosmological constant on the right hand side, since it includes the finite renormalized part. It is $\Lambda_{eff}=\frac14 \langle \tau^\mu{}_\mu \rangle +\Delta \Lambda$, after the constant contribution is extracted from the equations by averaging over the whole spacetime (denoted by $\langle \ldots \rangle$).
The modifications of the global sector of the theory, which can be written using a completely local action, lead to dynamical global constraints 
that ensure that all loop corrections precisely cancel from $\Lambda_{eff}$, once it is fixed by observations at a desired level of the loop expansion.

However, radiative stability is just one aspect of the cosmological constant problem in QFT coupled to gravity. Another aspect that is physically 
well defined is the issue of contributions from phase transitions \cite{wein,dreitlein,linde,veltman}. The problem is that phase transitions in gravitating QFT generically change the finite part of the vacuum energy by terms of the order of ${\cal O}({\cal M}^4)$, where ${\cal M}$ is the scale of the phase transition. For example, since QCD undergoes a confining phase transition at a scale $\sim  {\rm GeV}$, one expects that the vacuum energies of the QCD vacuum before and after the phase transition will differ by ${\cal O}({\rm GeV}^4) \simeq 10^{46} M_{Pl}^2 H_0^2$, where $H_0$ is the Hubble scale now. This is over 40 orders of magnitude greater than the critical energy density of the universe now. This begs the question, how did the theory know to pick the vacuum energy of the QCD vacuum {\it before} the phase transition just right, so that it cancels the contribution from the QCD phase transition to 46 decimal orders! Similar situation occurs with the electroweak phase transitions, GUT phase transitions, inflationary dynamics etc. 

In \cite{KP1,KP2} the problem of phase transitions and their contributions to vacuum energy was addressed qualitatively. The framework was the simple vacuum energy sequestering with global constraints. The main assumption made in that analysis was that the vacuum transition occurs instantaneosly over spacelike surfaces. As a consequence, it was found that the vacuum energy contribution from the transition is suppressed by the large volume of the universe after the transition. In a way, what helped suppress the vacuum energy difference was the fact that the universe spent a relatively short time in the false vacuum. On the other hand, while this picture of the post-transition geometry might be reasonable at late times after the transition, early on it does not capture the fact that vacuum transitions affect the geometry locally, via bubble nucleation and their subsequent evolution. Further, in the original sequestering scenario, the global constraints needed the spacetime volume of the universe to be finite, and the universe to collapse, and so studying local dynamics of phase transitions, and ultimately the geometric picture of a cosmic multiverse was difficult \cite{KP1,KP2}. However the manifestly local formulation of \cite{KPSZ} allows us to develop a fully local description which extends the dynamics of phase transitions and bubble formation from GR to the sequestering theory.

In this paper we study the effects of phase transitions in the matter sector on the geometry of the universe with the vacuum sequestering dynamics enforced by the $4$-form gauge sectors as in \cite{KPSZ}. We determine the post-transition geometry for the bubble dynamics using a set of adapted Israel junction conditions, and analyze the possible transitions in the limit of maximally symmetric geometries which describe the possible vacua of the theory. We work in the single bubble limit because in a large universe it suffices to understand the general dynamics when the nucleation rate is small (ie under control in EFT). This is similar to the situation in GR \cite{Coleman}. We find that requiring the absence of catastrophic instabilities imposes constraints on the functions $\sigma(\Lambda/\mu^4)$, $\hat \sigma(\kappa^2/M_{Pl}^2)$, requiring their logarithmic derivatives to be positive and sufficiently large, respectively. Since these functions are largely unconstrained by perturbative dynamics, these conditions can be readily met.
Once they are satisfied, the nucleation and evolution processes follow closely those of the standard GR, with no surprises. This shows that the sequestering mechanism remains consistent with the description of a universe with many phase transitions. Importantly, we compute the effect of the vacuum energy difference of the states before and after the transition on the geometry inside the bubbles of true vacuum. It is controlled by a ratio of the spacetime volumes before and after the transition, which are individually divergent. We regulate them by a time-reversal symmetric cutoff, picked out by the covariant junction conditions. In the limit when the cutoff surfaces approach the infinitely inflated past of the de Sitter (dS) parent, and the infinitely inflated future of the descendent, we find that the vacuum energy corrections from the phase transitions inside the descendant bubbles are completely suppressed. This shows that the corrections to vacuum energy from phase transitions in the regions described by the true dS (or Minkowski) vacua at late times are negligible, as it should be to fit the real world. 

The paper is organized as follows. In the next section we review the local vacuum energy sequestering of \cite{KPSZ}. In section 3, we study the process of vacuum energy transition, giving the description of the bubble formation and evolution in the thin wall limit, following the work of \cite{Coleman} for GR. In section 4, we study in more detail how bubbles evolve and consider the limits when they go to their maximal extent. We also compute the corrections to the vacuum energy generated by the jump of the potential energy density in field theory, and find that it is suppressed inside the bubbles of dS. We summarize in section 5. In the appendices we give additional technical details.

\section{Manifestly local vacuum sequestering: a review}

The local action for sequestering is given by \cite{KPSZ}
\begin{multline}\label{localaction1}
S= \int d^4 x \sqrt{g} \left[ \frac{\kappa^{2}(x)}{2} R  - \Lambda(x) - {\cal L}_m( g^{\mu\nu} , \Phi) \right] + \\
\int dx^{\mu}dx^{\nu} \ldots 
\left[\sigma\left(\frac{ \Lambda}{ \mu^4}\right)\frac{F_{\mu\nu\lambda\sigma} }{4!} +\hat \sigma\left(\frac{ \kappa^{2}}{ M_{Pl}^2}\right)\frac{\hat{F}_{\mu\nu\lambda\sigma}}{4!}  \right] \, .
\end{multline}
The metric $g_{\mu\nu}$ has corresponding Ricci scalar, $R$, $F_{\mu\nu\lambda\sigma} = 4\partial_{[\mu}A_{\nu\lambda\sigma]}$ and $\hat{F}_{\mu\nu\lambda\sigma} = 4\partial_{[\mu}\hat{A}_{\nu\lambda\sigma]}$ are a pair of $4$-forms and $\kappa(x)$ and $\Lambda(x)$ are scalar fields, and $\sigma$ and $\hat{\sigma}$ are  smooth functions whose arguments have been normalized with respect to the field theory cut off $\mu$ and the gravitational cut off $M_{Pl}$ respectively.  The functions are almost arbitrary: $\sigma$ must not be a logarithm \cite{KP2}, and $\sigma$ and $\hat \sigma$ must not both be linear\footnote{When both functions are linear both $4$-forms are completely specified by the geometry, which translates into a global constraint on the boundary data. There is no longer a one to one map between boundary data and observables: the former is one degree of freedom short on account of the constraint, and so the latter must relate to the fine-tuning of at least one parameter in the theory.}. The field theory sector is coupled to the metric minimally in the usual way. The second line in (\ref{localaction1}) is a purely non gravitating, topological sector by virtue of the absence of the metric. The pair of $4$-forms play the role of the covariant measure and their variation fixes $\kappa(x)$ and $\Lambda(x)$ to be integration constants on shell. However these scalars are fields which vary off shell, and their variation and selection of background values by the ensuing field equations are the origin of the constraint on the spacetime average of the Ricci scalar. On shell, they are constant because of their coupling to the $4$-forms, whose gauge symmetries completely remove the local degrees of freedom. These couplings of $\kappa(x)$ and $\Lambda(x)$ to the $4$-forms as well as to the gravitational sector ensure the constraint on $\langle R \rangle$ which yields the equation for the bare counterterm for the cosmological constant that guarantees cancellation of the loop corrections. Specifically the local field equations are   
\beastar
\kappa^{2} G^\mu{}_\nu = (\nabla^\mu \nabla_\nu-\delta^\mu{}_\nu \nabla^{2} )\kappa^{2} + T^\mu{}_\nu-\Lambda(x) \delta^\mu{}_\nu \, , \label{graveq} \\
\frac{\sigma'}{\mu^4} F_{\mu\nu\lambda\rho} = \sqrt{g} \epsilon_{\mu\nu\lambda\rho}, ~~
\frac{\hat \sigma'}{M_{Pl}^2}\hat F_{\mu\nu\lambda\rho} =-\frac{1}{2} R \sqrt{g} \epsilon_{\mu\nu\lambda\rho} \, , \label{Fs} \\
\frac{\sigma'}{\mu^{4}} \partial_{\mu}\Lambda = 0, ~~ \frac{\hat \sigma'}{M_{Pl}^{2}} \partial_{\mu}\kappa^{2} = 0 \label{sigma} \, ,
\eeastar
where $T_{\mu\nu} = \frac{2}{\sqrt{g}}\frac{\delta}{\delta g^{\mu\nu}}\int d^{4}x \sqrt{g}\mathcal{L}_{m}(g^{\mu\nu}, \Phi)$ is the matter stress energy tensor. Taking the trace of the gravity equation and averaging over spacetime fixes the value of the classical counter term $\Lambda$ (fixed to be constant by the $F_{4}$ equation of motion) as a function of $\langle T^\alpha{}_\alpha \rangle$ and $\langle R \rangle$ where $T^\alpha{}_\alpha = g^{\mu\nu} T_{\mu\nu}$ is the trace of the matter stress energy tensor. One can then eliminate the dependence of $\langle R \rangle$ using the integrated equations for the two $4$-forms such that
\be
\Lambda = \frac{1}{4}\langle T^\alpha{}_\alpha \rangle + \Delta \Lambda
\ee
with 
\be
\Delta \Lambda=\frac14 \kappa^{2}  \langle R \rangle=-\frac{\mu^4}{2} \frac{\kappa^{2}  \hat \sigma'}{M_{Pl}^2 \sigma'}\frac{\int \hat F_4}{\int F_4} \, . \label{DelLam2}
\ee
Inserting this expression into the gravity equation yields
\be
\kappa^{2} G^\mu{}_\nu = T^\mu{}_\nu - \frac{1}{4} \delta^\mu{}_\nu \langle T^\alpha{}_\alpha \rangle - \Delta \Lambda  \delta^\mu{}_\nu \, .
\ee
So as claimed above, the vacuum energy contributions to the energy momentum tensor ($T^\mu{}_\nu = -\delta^\mu{}_\nu V_{vac}$) will drop out and not source curvature. 
We refer the reader to \cite{KPSZ} for more details.

In the presence of a boundary, the action (\ref{localaction1}) must be supplemented with extra terms in order to have a well defined variational principle.  These are the analogue of the Gibbons-Hawking term in General Relativity \cite{GH} and they will affect the computation of tunnelling rates in the next section.  To this end, we supplement the action with the following  boundary term
\be
\int d^3 x  \sqrt{h} \kappa^2(x) K \, ,
\ee
where $K$ is the trace of the extrinsic curvature on the boundary, and $h_{ij}$ the induced metric. Variation of the action with respect to the metric and $\kappa$ now yields a boundary term \cite{vish}
\be \label{boundaryterm}
\int d^3 x   \frac12 \sqrt{h} \left[-\kappa^2 (K^{ij}-K h^{ij})\right] \delta h_{ij}+\sqrt{h} K\delta \kappa^2 \, ,
\ee
where we have used $n^a \partial_a \kappa^2=0$ which follows from the bulk equations of motion. For the action to be stationary under such a variation one normally imposes Dirichlet boundary conditions on both the metric and $\kappa$.  However, we do not do so here. Dirichlet boundary conditions on either $\Lambda$ and $\kappa$ would interfere with the global constraints that arise from bulk variations, and are crucial to the sequestering mechanism. Instead we imagine imposing Neumann boundary conditions on the two scalars, which physically means that there is no momentum loss from the scalar through the boundary. However, since the scalars are constant on shell, the 
Neumann boundary conditions  are really redundant since they  are automatically satisfied by the solutions of the field equations.

For the action to be stationary under all field variations, we impose an alternative boundary condition on the metric,
\be \label{bc}
\delta h_{ij}=-\frac{\delta \kappa^2}{\kappa^2} h_{ij} \, ,
\ee
and Dirichlet boundary conditions on the $3$-forms.  The choice (\ref{bc}) is actually equivalent to a Dirichlet boundary condition on the {\it Einstein frame} metric. In any event, it now follows that the variation of the full action does indeed vanish on-shell, as required by  a well defined action and variational principle.

\section{Inclusion of  sequestering: materialization of the bubble}

Let us now look at the case when the matter sector has two vacua with  different vacuum energy.  We will treat the system semiclassically, and allow for the matter sector to tunnel quantum mechanically between vacua. To derive the tunnelling rates we must compute the bounce, a solution to the Euclidean field equations\footnote{The Wick rotation to Euclidean signature is $
t \rightarrow -i t_{E},  S \rightarrow i S_{E}$.} that interpolates between the two vacua.
In the limit where the difference in the energy density between vacua is small compared with the height of the barrier separating them, we can describe the transition region using a thin wall approximation \cite{ColemanDeLuccia}.
Generically one expects vacuum decay to be dominated by Euclidean configurations that are $O(4)$ invariant \cite{O4,O3O2}. We will therefore also take a bounce geometry with  rotational invariance such that $ds^{2} = dr^{2} + \rho^{2}(r) d\chi^{2}$ where $d \chi^{2}=\gamma_{ij}dx^idx^j$ is the unit 3-sphere in Euclidean signature.

In a neighborhood of the wall, we can set up a coordinate system such that the wall is centered at $r=0$ with $r>0$ corresponding to the exterior of a bubble (which we will denote ${\cal M}_{+}$) and $r<0$ the interior (which we denote  ${\cal M}_{-}$). Due to the rotational invariance, all fields are now only dependent on the radial coordinate $r$.  In particular, we have the following non-zero components for the $3$-forms
\be
A_{ijk}=A(r)\sqrt{\gamma}\epsilon_{ijk}, \qquad \hat A_{ijk}=\hat A(r)\sqrt{\gamma}\epsilon_{ijk} \, .
\ee
$\Lambda$ and $\kappa$ are constant on shell, while the remaining field equations reduce to the following set of ordinary differential equations
\begin{eqnarray}
3 \kappa^2 \left( \frac{\rho'^2}{\rho^2}-\frac{1}{\rho^2} \right) &=& -(\Lambda+V(r)) \, , \label{rhodot}\\
\kappa^2  \left( \frac{\rho'^2}{\rho^2}-\frac{1}{\rho^2}+2\frac{\rho''}{\rho} \right) &=& -(\Lambda+V(r)+\sigma_w \delta (r)) \, , \label{rhoddot} \\
\frac{\sigma'}{\mu^4} A'(r) &=& \rho^3 \, , \label{A}\\
 \frac{\hat \sigma'}{M_{Pl}^2} \hat A'(r) &=&-3\left(\frac{1}{\rho^{2}} - \frac{\rho'^{2}}{\rho^{2}} - \frac{\rho''}{\rho}\right) \rho^3 \, . \qquad \label{hatA}
\end{eqnarray}
Here we have taken the thin wall limit, modelling  the contributions from  vacuum energy as a step function
\be
V(r)=\begin{cases} V_+ & r>0 \\ V_- & r<0 \end{cases}
\ee
and the wall as a delta function weighted by its  tension $\sigma_w$.  Away from the bubble wall, we see that
\be \label{rho}
\rho(r) = \frac{1}{q} \sin q(r_{0} + \epsilon r)
\ee
where $\epsilon = \pm 1$, and 
$$q^{2} = \frac{\Lambda +V}{3\kappa^2}$$
 represents the local value of the vacuum curvature.   The expression (\ref{rho}) holds for $q$ real or pure imaginary, and in the limit that $q \to 0$, corresponding to sections of the sphere, hyperboloid and plane respectively\footnote{When we Wick rotate back to Lorentzian signature in the next section, these will correspond to sections of dS, anti deSitter (AdS) and Minkowski space respectively}. $r_0$ is an integration constant that can take different values on either side of the wall.   The solutions for the $3$-forms are
\begin{eqnarray}
A(r) &=& A_0+\frac{\mu^4}{\sigma'} \int_0^r \rho^3 dr \, , \\
\hat A(r) &=& \hat A_0-6\frac{M_{Pl}^2}{ \hat \sigma'} \int_0^r q^2 \rho^3 dr \, ,
\end{eqnarray}
where again, the integration constants $A_0$ and $\hat A_0$ can, in principle, differ on either side of the wall.

We must now impose  matching conditions across the bubble wall. These normally take the form of continuity conditions on the dynamical fields, and junction conditions\footnote{In General Relativity these are usually referred to as the Israel junction conditions \cite{Israel}} on their  normal derivatives. Here the situation turns out to be slightly more subtle on account of the fact that not all fields are dynamical. We can extract the appropriate matching conditions by simply integrating the field equations (\ref{rhodot}) to (\ref{hatA}) across the bubble wall. Equations (\ref{rhodot}) and (\ref{A})  respectively result in continuity conditions on the radius of the 3-sphere $\rho$, and the $3$-form, $A$,  yielding
\be
\frac{1}{q_+} \sin q_+ r^+_{0} =\frac{1}{q_-} \sin q_- r^-_{0} , \qquad A_0^+=A_0^- \, ,
\ee
where the labels $\pm$ correspond to ${\cal M}_\pm$.  In contrast, integrating equations (\ref{rhoddot}) and (\ref{hatA}) across the wall, we find the following discontinuities supported by the delta function source
\be \label{discs}
2\kappa^2  \frac{\Delta \rho'}{\rho_0}=-\sigma_w, \qquad \frac{\hat \sigma'}{M_{Pl}^2} \Delta \hat A=3\rho_0^2 \Delta \rho' \, ,
\ee
where $\rho_0=\rho(0)$ and $\Delta Q=Q(0^+)-Q(0^-)$ denotes the jump across the wall.  The first discontinuity (ie the jump in $\rho'$) is familiar: it represents the jump in extrinsic curvature across the bubble wall as per the Israel junction conditions \cite{Israel}.  However, we see that we also have a second  discontinuity, from which we can infer the following jump in the $3$-form, $\hat A$, 
\be
\hat A_0^+-\hat A_0^-=-\frac{3 M_{Pl}^2}{2\kappa^2 \hat \sigma'} \rho_0^3 \sigma_w \, .
\ee
This discontinuity in the $3$-form  follows from the fact that the corresponding $4$-form field strength is sourced by the curvature, which in turn gets sourced by the delta function at the wall.  In a physically resolved configuration with a wall of finite thickness, clearly all fields would be continuous (at least up to a gauge transformation). That field $\hat A$, however, would vary nontrivially -- with the variation becoming a sharp jump in the thin wall limit -- occurs because a wall with a finite tension is charged under it. This is because $\hat A$ couples to (vacuum) energy, and so tensional walls behave as membranes carrying its charge.

If we demand that the  wall is supported by physically realistic matter, we require that the  tension of the wall be non-negative ($\sigma_w \geq 0$). This places an important constraint on the allowed configurations following from (\ref{discs}), 
\be
\Delta \rho' = \Delta(\epsilon \cos qr_{0}) \leq 0 \, .
\ee
An identical constraint is obtained in GR. We can place further constraints by taking into account  the impact of tunnelling rates, which we will now compute. 

\begin{figure}[thb] 
  \centering
\includegraphics[height=1in]{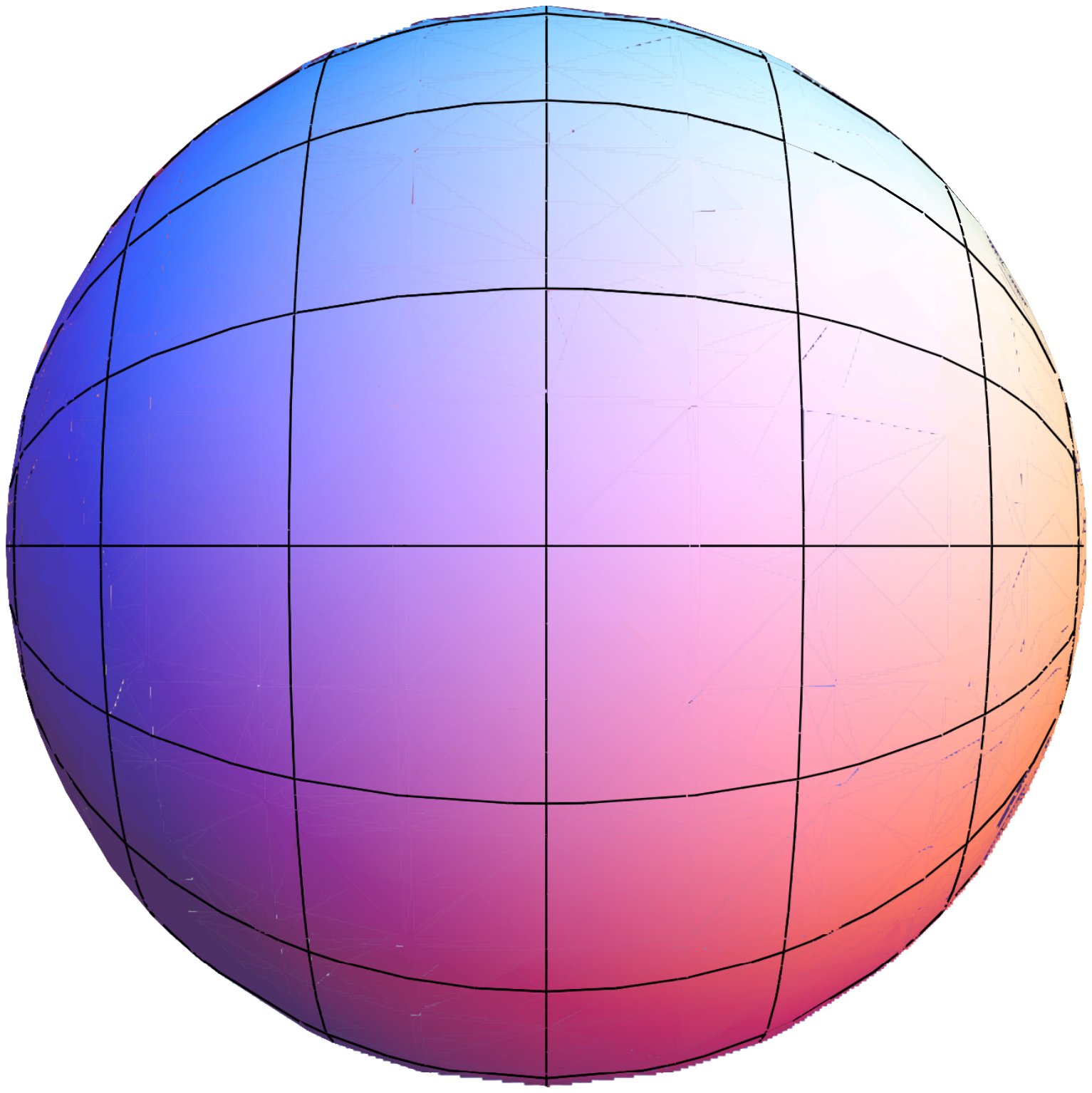}
\includegraphics[height=1in]{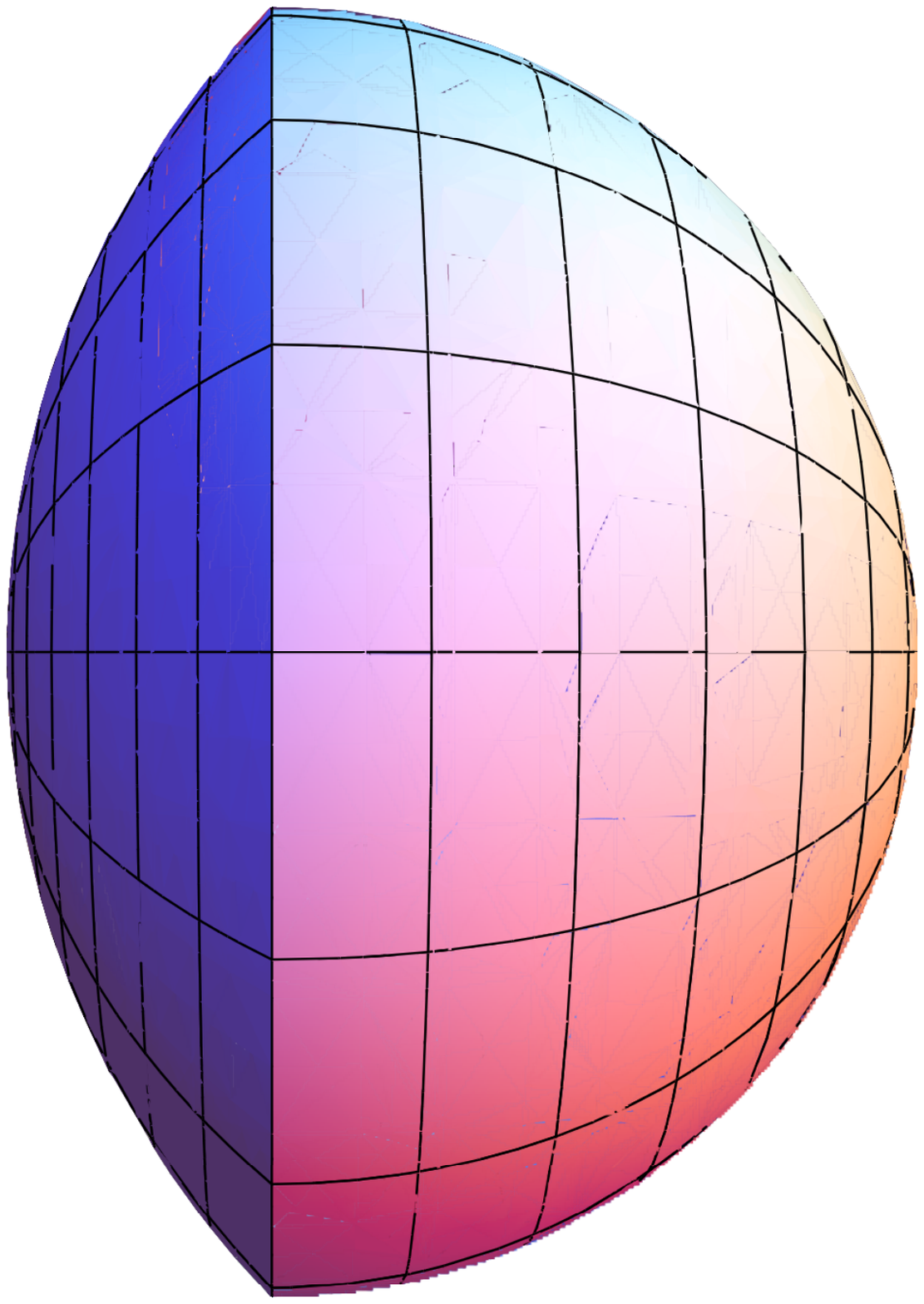} 
    \caption{Pictoral representations of an initial Euclidean vacuum with positive curvature (left) and the corresponding bounce describing tunnelling to a vacuum with lower positive curvature (right). }
    \label{bounce}
\end{figure}

In semiclassical theory of vacuum decay in the presence of gravity,  tunnelling rates describing transition between  vacua  are given by \cite{Coleman,CallanColeman,ColemanDeLuccia}
\be
\frac{\Gamma}{V} \sim \exp ^{-B/ \hbar} \, ,
\ee
where
\be
B=S_{E}^{bounce}-S_E^{\infty} \, .
\ee
Here $S_{E}^{bounce}$ is the Euclidean action
evaluated on the bounce solution  and $S_E^{\infty}$ is the Euclidean action for the initial vacuum solution.  The bounce solutions  are just the Euclidean bubble configurations described above. The entire solution is covered by the  coordinates given in the neighborhood of the wall, with $r$ ranging from its minimum value in the interior, $r_{min}^-$ and its maximum value in the exterior $r_{max}^+$, where
 \be
 r_{min}=\begin{cases} -r_0 \, , & \epsilon=+1 \, , \\
  r_0-\frac{\pi}{q} \, , & \epsilon =-1, ~q^2 >0 \, , \\
  -\infty \, , &  \epsilon =-1, ~q^2 \leq 0 \, , \end{cases} 
 \ee
 \be
 r_{max}=\begin{cases}
  \frac{\pi}{q}-r_0 \, , & \epsilon =+1, ~q^2>0  \, , \\
  \infty \, , &  \epsilon =+1, ~q^2 \leq 0 \, , \\
   r_0 \, , & \epsilon=-1 \, .\\
 \end{cases} 
 \ee
The initial vacuum solution is taken to range from $r_{min}$ and $r_{max}$, but without any jump in the curvature. This solution  coincides with the bounce in the exterior of the bubble.  To gain an intuitive picture for these configurations consider  tunnelling between vacua of positive curvature, as in Fig. \ref{bounce}: the initial vacuum is just a Euclidean sphere, whilst the bounce is two such spheres of different radii, cut and pasted together across the bubble wall.
 
For sequestering the on shell Euclidean action is closely related  to the one in GR, differing only by the flux terms, such that
 \be \label{B}
 B=B_{GR}-\sigma \Delta c -\hat \sigma \Delta \hat c \, .
 \ee
Here $B_{GR}=-2\kappa^2 \Omega_3 \Delta \left[\frac{1}{ q^2} [\rho'^3]^{0}_{r_{min}}\right]+\sigma_w \Omega_3 \rho_0^3$,  represents the tunnelling exponent computed in GR for the same geometrical configuration, and $\Omega_3$ is the volume of the unit $3$-sphere. In GR,  $\Lambda$ is a fixed parameter that has to be chosen by hand, in contrast to the sequestering scenario. Although the asymptotic structure of the $3$-forms agree for the bounce and the initial vacuum, the net flux does not, on account of the fact that the total volumes can and do differ. Because of this, we have non-trivial values for
 \be
 \Delta c=\int_\text{bounce} F_4-\int_\text{initial~vac} F_4=-\frac{\mu^4}{\sigma'}\Omega_3 \Delta \left[\int_{r_{min}}^0 dr \rho^3 \right] \, ,
 \ee
 and
 \begin{multline}
 \Delta \hat c=\int_\text{bounce} \hat F_4-\int_\text{initial~vac} \hat F_4 \, , \\
 =\frac{M_{Pl}^2}{2 \hat \sigma'}\Omega_3 \left(-3 \frac{\sigma_w}{\kappa^2} \rho_0^3+12 \Delta \left[ \int_{r_{min}}^0 dr  q^2 \rho^3 \right]\right) \, .
 \end{multline}
One can easily show that
 \be
 \int_{r_{min}}^0 dr \rho^3 =-\frac{1}{3 q^4} [\rho'(3-\rho'^2)]^0_{r_{min}} \, .
 \ee
Now divergences in (\ref{B}) can occur when $r_{min}=-\infty$, or in other words, when $\epsilon=-1$, and we have $q^2 \leq 0$ (ie planar or hyperbolic geometries).  These are important since they can lead to infinitely suppressed ($B \to +\infty$) or infinitely enhanced ($B \to -\infty$) tunnelling rates. Provided we have non-negative tension in the wall, the only configurations we need to worry about are those with $\epsilon_+=-1$ and $q_+^2 \leq 0$. These are wormhole configurations -  in the Lorentzian picture, the asymptotic region of AdS or Minkowski space tunnels to a new vacuum. However in GR, for these $B_{GR} \to +\infty$, and so they are infinitely suppressed.  This need not be true in sequestering. To see this, note that the divergent contribution to the tunnelling exponent goes as
\be
\sim \frac{\Omega_3}{8}\left[ \frac{2\kappa^2}{|q|_+^2}\left( 1+\frac{M_{Pl}^2 \hat \sigma}{\kappa^2 \hat \sigma'}\right)+\frac{\mu^4}{3 |q|_+^4}\frac{  \sigma}{\sigma' }\right]e^{-3|q|_+ (r_{min}^+-r_0^+)} \, . \label{whrate}
\ee
Depending on the form of $\sigma$ and $\hat \sigma$ this may diverge to either $+\infty$, or even $-\infty$. The latter would correspond to a catastrophic vacuum instability -- the semiclassical approximation breaks down completely, and tunnelling rates are unsuppressed.  However, such pathologies are avoided with a judicious choice of $\sigma$ and $\hat \sigma$, which as we noted above are largely unconstrained by perturbative physics. This situation is in fact really similar to what happens in GR when one adds higher derivative boundary terms, which can completely alter the Euclidean actions \cite{lennyandco}. So as in those case, we will simply assume that a choice of boundary contributions has been made which precludes instabilities and ignore these wormhole configurations.  With this, the net result is that the spectrum of allowed configurations is identical to those in GR, as seen from  table \ref{table:configs2}.
\begin{table*} 
\centering
\begin{tabular}{| c|| c |c| c| c|}
\hline
&$\text{S}_{+}-\text{S}_{-}$&$\text{S}_{+}-\text{H}_{-} $ &$\text{H}_{+}-\text{S}_{-} $&$\text{H}_{+}-\text{H}_{-} $ \\ 
\hline\hline
$\epsilon_{\pm} = 1$ & $(qr_{0})_{+} \geq (qr_{0})_{-} $ & allowed & not allowed & $|q|_{+} \leq |q|_{-} $  \\  \hline
$\epsilon_{\pm} = -1$ & $(qr_{0})_{+} \leq (qr_{0})_{-} $ & not allowed & infinitely suppresed & infinitely suppresed   \\ \hline
$\epsilon_{+} = 1, \epsilon_{-} = -1$ & $\langle qr_{0} \rangle \in [\pi/2, \pi]$ & not allowed & not allowed & not allowed \\ \hline
$\epsilon_{+} = -1, \epsilon_{-} = 1$ & $\langle qr_{0} \rangle \in [0, \pi/2]$ & allowed & infinitely suppresed  &infinitely suppresed \\
\hline
\end{tabular}  
\caption{  \label{table:configs2} Summary of allowed configurations, taking into account the constraint on the  tension {\it and} assuming wormhole nucleation  suppressed in Minkowski and AdS. $S$ denotes the sphere, $H$ the hyperboloid, and planar limits can be extracted from the table simply by taking $q_{+}$ or $q_{-}$ to vanish. } 
\end{table*} 

For all of the remaining configurations with walls of non-negative tension, it turns out that
\be
\rho'(r_{min})=1, \qquad -1 \leq \rho'(0^+) \leq \rho'(0^-) \, .
\ee
One can next show
\be
B_{GR}=2\Omega_3 \kappa^2 \rho_0^2 \Delta \left[ \frac{1}{1+\rho'(0)} \right]  \geq 0 \, ,
\ee
while the flux contributions are
\be
-\sigma \Delta c= \Omega_3\frac{\mu^4 \rho_0^4}{3} \frac{\sigma}{\sigma'}  \Delta \left[ \frac{1}{1+\rho'(0)} +\left( \frac{1}{1+\rho'(0)} \right)^2 \right] \, ,
\ee
\be
-\hat \sigma \Delta \hat c=-\Omega_3M_{Pl}^2\rho_0^2  \frac{\hat \sigma}{\hat \sigma'} \Delta \left[ \rho'(0)+\frac{4}{1+\rho'(0)} \right] \, .
\ee
Although tunnelling is guaranteed to be suppressed in GR, once again,  this is not {\it a priori} guaranteed for sequestering because generically the sign of the flux contributions is not fixed. As above we must make a judicious choice of $\sigma$ and $\sigma'$ in order to avoid a breakdown of the semiclassical description. If we also take into account the wormhole tunnelling rates described by (\ref{whrate}), a sufficient condition for all tunnelling rates to remain suppressed is given by
\be
\frac{\kappa^2 \hat \sigma'}{M_{Pl}^2 \hat \sigma} >2, \qquad  \frac{\sigma'}{ \mu^4 \sigma} >0 \label{cond} \, .
\ee
The conditions are very important since they place constraints on the theory required by the absence of rapid instabilities.
These inequalities can be satisfied, for instance, using a growing exponential for $\sigma$, and a monomial for $\hat \sigma$.  When the conditions (\ref{cond}) hold we find that sequestering admits exactly the same transitions as GR, albeit with slightly modified rates. Tunnelling between vacua with positive curvature can proceed in either direction, although the transition rate for tunnelling upwards is significantly suppressed as in GR. If we wish to consider  other values of the curvature as well (ie  zero and negative) then tunnelling will always proceed towards vacua of lower curvature, with the reverse process proving impossible.  The interior of bubbles of zero or negative curvature can only contain the centre of the geometry as opposed to the asymptotic region ($\epsilon_-=+1$). Similarly, the  exterior regions of zero or negative curvature must always contain the asymptotics ($\epsilon_+=+1$).

To get some intuition as to the origin of the bounds (\ref{cond}), note that the Euclidean action for a sphere of radius $1/q$ is given by 
\be \label{SEGR}
S_{E}^{GR}=-6 \pi^2 \frac{M_{Pl}^2}{q^2} \, ,
\ee
for GR, and 
\be \label{SEseq}
S_{E}^{seq}=-6 \pi^2 \frac{M_{Pl}^2}{q^2} \left(\frac{\kappa^2}{M_{Pl}^2}-2 \frac{\hat \sigma}{\hat \sigma'}\right)-2 \pi^2 \frac{\mu^4}{q^4} \frac{ \sigma}{ \sigma'} \, ,
\ee
for sequestering. When the bounds (\ref{cond}) are saturated, we see that the Euclidean action for a sphere of any radius vanishes in sequestering. This represents a critical point where the qualitative behavior changes: when the bounds are satisfied, increasing the radius of the sphere always lowers the Euclidean action, just as it does in GR. When the bounds are not satisfied this is no longer true.  The bounds guarantee qualitatively similar behavior to GR, favoring transitions that lower the dS curvature.  Of course, the sequestering corrections can and do alter the rate at which the Euclidean action decreases with the curvature of the sphere, and this will affect tunnelling rates. 

Let us for simplicity examine the same two special, but illuminating cases as \cite{ColemanDeLuccia}.
The first is decay  from positive  into zero curvature ($q^2 \to 0$).  We then have $\rho'(0^-)=1$ and $\rho'(0^+) \in [-1,1]$. The tunnelling exponent is %
\be
B=B_{GR} \left[ 1+\frac{\mu^4}{12 q^2 \kappa^2} \frac{\sigma}{\sigma'}s(8-3s)-\frac{M_{PL}^2}{\kappa^2}\frac{\hat \sigma}{\hat \sigma'}s \right] \, ,
\ee
where $B_{GR}=\Omega_3 \frac{\kappa^2}{q^2} s^2$ and 
\be
s=1-\rho'(0^+)=\frac{\sigma_w^2}{2\kappa^4 q^2}\left( \frac{1}{1+\sigma_w^2/4 \kappa^4 q^2}\right) \, .
\ee
For this process,  sequestering can either enhance or suppress tunnelling relative to GR, depending on the choice of $\sigma$ and $\hat \sigma$, and the size of the bubble. As with GR, the dominant processes correspond to nucleation of small bubbles, with $0 \leq s \ll 1$, for which
\be
B \approx B_{GR}\left[1+\frac{2\mu^4}{3 q^2 \kappa^2} \frac{\sigma}{\sigma'}s-\frac{M_{PL}^2}{\kappa^2}\frac{\hat \sigma}{\hat \sigma'}s \right] \, .
\ee
Assuming the conditions (\ref{cond}) hold, we see that for large jumps in curvature tunnelling is enhanced by the sequester, as the hatted fluxes win out, whereas for small jumps it its suppressed, since the unhatted fluxes win out in this case. 
We can  understand this intuitively by once again studying the Euclidean actions for a sphere (\ref{SEGR}, \ref{SEseq}). For high curvatures, the first term in (\ref{SEseq}) dominates and we see that sequestering reduces the rate at which the action decreases, in comparison to GR. This should make tunnelling easier  as there is less suppression.   At low curvatures the second term dominates and sequestering enhances the rate at which the action decreases, making tunnelling more difficult.

Next we consider decay from zero to negative curvature ($0 \to -|q|^2$). We have $\rho'(0^+)=1$ and  $\rho'(0^-) \geq 1$, so that the tunnelling exponent is given by
\be
B=B_{GR} \left[ 1-\frac{\mu^4}{12 |q|^2 \kappa^2} \frac{\sigma}{\sigma'}s(8-3s)-\frac{M_{PL}^2}{\kappa^2}\frac{\hat \sigma}{\hat \sigma'}s \right] \, ,
\ee
where $B_{GR}=\Omega_3 \frac{\kappa^2}{|q|^2} s^2$ and 
\be
s=1-\rho'(0^-)=-\frac{\sigma_w^2}{2\kappa^4 |q|^2}\left( \frac{1}{1-\sigma_w^2/4 \kappa^4 |q|^2}\right) \, .
\ee
As emphasized in \cite{ColemanDeLuccia}, there are no sensible solutions with $|q|^2< \sigma_w^2/4\kappa^4$. This is already well understood: for a transition to occur the energy stored in the wall should compensate for the energy deficit inside the bubble, and in AdS space the  bubble wall simply cannot get big enough for this to happen. Indeed, this result was recently shown to extend beyond the thin wall limit \cite{Masoumi}.
Again as in GR, the dominant processes correspond to nucleation of small bubbles, with $0 <-s \ll 1$. For the decay of Minkowski into AdS, this limit yields a tunnelling exponent
\be
B \approx B_{GR}\left[1-\frac{2\mu^4}{3 |q|^2 \kappa^2} \frac{\sigma}{\sigma'}s-\frac{M_{PL}^2}{\kappa^2}\frac{\hat \sigma}{\hat \sigma'}s \right] \, .
\ee
Recall from \cite{ColemanDeLuccia} that in GR gravitational effects are seen to help stabilize the false Minkowski vacuum.  Assuming the conditions (\ref{cond}) hold, 
we see  that sequestering enhances this effect even further.

\section{Inclusion of sequestering: growth of the bubble}

Once the bubble has materialized we can track its subsequent evolution by Wick rotating the bounce solution back to Lorentzian signature. In  a neighborhood of the bubble wall, the geometry is described by the metric
\be
ds^2 =dr^2 +\rho(r)^2 (- d \tau^{2} + \cosh^{2} \tau d\Omega^{2}_{2})  \, ,  \label{colecoords}
\ee
where
\be
\rho(r)=\begin{cases} \frac{1}{q_+} \sin q_+ (\epsilon_+r+r^+_{0}) \, , & r > 0 \, ,  \\ \frac{1}{q_-} \sin q_- (\epsilon_- r+r^-_{0}) \, , & r<0 \, , \end{cases}
\ee
with the wall itself at $r=0$.  The interior and exterior correspond to sections of maximally symmetric spacetime, ie Minkowski ($q^2=0$), dS ($q^2>0$), or AdS ($q^2<0$), although it will be important to realize that these coordinates do not cover the entire space in Lorentzian signature. $\tau=0$ is a special point since it corresponds to a minimal spacelike surface with vanishing extrinsic curvature. This represents a stationary point in the geometry where one may consistently Wick rotate to Euclidean time and connect to the bounce solution. It follows that $\tau=0$ corresponds to the nucleation time: before this there is no wall and  the entire solution lies in the initial vacuum (labelled with a ``+"). Afterwards we have an expanding bubble separating two distinct vacua (labelled with ``+" and ``-"). We saw in the previous section that the allowed configurations match those in GR. In others words:  (i) we may tunnel between dS vacua in either direction, although the process which increases the dS curvature is suppressed; (ii) all other allowed tunnelling processes serve to reduce the curvature; (iii) if the exterior of a bubble is Minkowski or AdS, it must include the asymptotic region; and (iv) if the interior of a bubble is Minkowski or AdS it must {\it not} include the asymptotic region (ie no wormholes)

Now the dynamics of the sequestering scenario is sensitive to the global structure of the solution, so Coleman's coordinate patch (\ref{colecoords}) is inadequate for a complete description. We therefore switch to global coordinates capable of covering the entire maximally symmetric spacetime. These are  described in detail in the appendix. We summarize the main features here. For  dS in global coordinates, the metric is given by
\be
ds^{2} = -dt^{2} + \frac{\cosh^{2} qt}{q^{2}}(d\theta^{2} + \sin^{2}\theta d\Omega_{2}^{2}) \, ,
\ee
where $t \in (-\infty, \infty )$ and $\theta \in [0, \pi]$. Tunnelling can occur at any of the minimal spacelike surfaces 
\be
\cos\theta=
\frac{\tanh qt}{\tanh \alpha} \, ,
\ee
parameterized by the constant $\alpha $.  Indeed, as shown in the appendix, one can locally map these surfaces to the $\tau=0$ surface in Coleman's coordinate patch. 

For the Minkowski vacuum, the globally defined metric is given by
\be
ds^2=-dt^2+du^2 +u^2 d\Omega_{2}^{2} \, ,
\ee
where $t \in (-\infty, \infty )$ and $u \in [0, \infty)$, with minimal spacelike surfaces occuring at $t=t_0$, constant. Again, it is shown in the appendix that one can locally map these surfaces to the tunnelling surface $\tau=0$ in Coleman's coordinates.

Finally, for AdS in global coordinates,
\be
ds^{2} = - \frac{\cosh^{2} |q|u}{|q|^{2}} dt^{2} + du^{2} + \frac{\sinh^{2}|q|u}{|q|^{2}} d\Omega_{2}^{2} \, ,
\ee
where $t,u \in (-\infty,\infty)$. Again, the minimal spacelike surfaces that map to $\tau=0$ correspond to $t=t_0$, constant.

Up until the  minimal spacelike surface the solution lies in a single vacuum described by the appropriate global coordinate system. Once the bubble has nucleated we have two distinct vacua separated by the bubble wall. As shown in the appendix, in the global coordinate systems the wall itself is located  at
\begin{eqnarray}
\textrm{ dS}&:& \cos q r_0=\cosh \alpha \cosh qt \cos \theta-\sinh \alpha \sinh qt \, , \qquad \nonumber \\
\textrm{flat}&:& r_0^2=u^2-(t-t_0)^2 \, , \qquad  \nonumber\\
\textrm{AdS}&:& \cosh |q| r_0=\cosh |q|u \cos(t-t_0)  \, . \qquad \nonumber
\end{eqnarray}
Outside the wall, the initial vacuum solution persists. Inside the wall a new vacuum appears, again described by an appropriate global coordinate system. The interior solution is cut off in space at the position of the wall, and in the past by a suitable minimal surface.

When the interior solution corresponds to a portion of AdS space there is an additional feature, already noted by Coleman et al in GR \cite{ColemanDeLuccia, AbbottColeman}.  The interior AdS geometry is unstable against gravitational collapse, with a curvature singularity on the surface
\be
\cosh |q|  u \cos (t-t_0)=-1 \, .
\ee

Armed with the global structure of our solutions we can proceed to compute  the integrated fluxes of the $4$-forms, 
which control the vacuum energy contributions in the bulk. As in the Euclidean case we denote these respectively as $c$ and $\hat{c}$, 
\bea
c = \int F_{(4)} = \frac{\mu^{4}}{\sigma^{'}} \int d^{4}x \sqrt{g} \, , \\
\hat{c} = \int \hat{F}_{(4)} = -\frac{M_{Pl}^{2}}{2 \hat{\sigma}^{'}} \int d^{4}x \sqrt{g} R \, ,
\eea
although generically the integrated Lorentzian fluxes will differ from their Euclidean counter parts.

For any given solution, the spacetime is split into the volume before bubble nucleation (denoted by $\V_b^+$), the volume of the exterior of the bubble after nucleation (denoted by $\V_a^+$), and the volume of the interior of the bubble after nucleation (denoted by $\V_a^-$). There is also the bubble wall, although its contribution will never be particularly significant. In any event, the integrated fluxes can be calculated explictly on a solution
\bea
c = \frac{\mu^{4}}{\sigma^{'}}(\V_{b}^{+} + \V_{a}^{+} + \V_{a}^{-}) \, , \label{ceqn}\\ 
\hat{c} = -\frac{6 M_{Pl}^{2}}{\hat{\sigma}^{'}}[(q^{2}\V_{b})^{+} + (q^{2}\V_{a})^{+} + (q^{2}\V_{a})^{-}] \, . \label{chateqn}
\eea
The total spacetime volume is  the sum $\V_{total} = \V_{b}^{+} + \V_{a}^{+} + \V_{a}^{-}$. 
Equation (\ref{ceqn}) and (\ref{chateqn}) can be rewritten as equations for the total cosmological constants in the two vacua as   
\bea
q_{+}^{2} = -\frac{\mu^{4}\hat{\sigma}^{'}}{6 M_{Pl}^{2} \sigma^{'}}\frac{\hat{c}}{c} + \frac{\Delta q^{2}}{1+{\cal I}} \, , \label{q+}\\
q_{-}^{2} = -\frac{\mu^{4}\hat{\sigma}^{'}}{6 M_{Pl}^{2} \sigma^{'}}\frac{\hat{c}}{c} - \frac{\Delta q^{2}}{1 + {\cal I}^{-1} } \label{q-} \, ,
\eea
where  
\be
\Delta q^{2} = q_{+}^{2} - q_{-}^{2}=\frac{\Delta V}{3\kappa^2} \, ,
\ee
and
\be
{\cal I} = \frac{\V_{b}^{+} + \V_{a}^{+}}{\V_{a}^{-}} \, .
\ee     
Consistent with the variational principle, we treat $c$ and $\hat c$ as fixed, and extract the above expressions (\ref{q+}, \ref{q-}),  for the local curvature. 

At this point, we can see that these equations are the key for understanding the effect of sequestering on the vacuum energy contributions by phase transitions, given by the jump in vacuum energy, $\Delta V$, which is induced by the transition inside each bubble. The influence of the jump on the geometry of the region where it is observed is controlled by  the ratios of volumes before and after bubble nucleation, in a way similar to the intuitive picture of volume controlled corrections in global sequester \cite{KP1,KP2}. Technically, we see that if ${\cal I} \gg 1$ we have that $q_{+}^{2} = -\frac{\mu^{4}\hat{\sigma}^{'}}{6 M_{Pl}^{2} \sigma^{'}}\frac{\hat{c}}{c}$, and so the exterior curvature  is completely insensitive to the jump in vacuum energy, and is given entirely by the residual cosmological constant.  In contrast, in this case the interior curvature would be strongly dependent on $\Delta V$. The reverse is true when ${\cal I} \ll 1$ and in general,  the ratio $\cal I$ determines which cosmological constant has the least sensitively to the local vacuum energy.

This behavior is of crucial importance for the physics of vacuum energy induced by phase transition. We can summarize it very simply:
\begin{itemize}
\item[]
{\it Vacuum energy is most efficiently sequestered in the vacuum that dominates the spacetime volume. The more it dominates, the more efficient the sequester in that region.} 
\end{itemize}
Again, as in \cite{KP1,KP2}, the vacuum energy was more efficiently sequestered at late times the earlier the transition, except now we have a fully local description, which is physically more realistic. We give the relevant spacetime volumes and their ratios in the appendix D. They formally diverge in the limit of infinite past and future, but the junction conditions which follow from covariance ensure that the divergence rates are the same. Thus the ratios are finite, and the regulator completely cancels. This cosmological version of the l'Hopital's theorem yields the following ratios for various geometries before and after the transition:
\bea
\I_{dS \to dS} &\sim & \frac{q_-}{q_+} \, , \\
\I_{dS \to M} &=& 0 \, , \\
\I_{dS \to AdS} &=& \infty \, , \\
\I_{M \to AdS} &=& \infty \, , \\
\I_{AdS \to AdS} &=& \infty \, ,
\eea
where $\I_{X \to Y}$ denotes tunnelling from X to Y, where X, Y are dS (dS), M (Minkowski) and AdS (AdS).

Consider the dS to dS transitions, for which transitions can occur in either direction. These are clearly of main phenomenological interest. For these initial and final geometries, the transitions from high to low curvature have $q_-<q_+$, and so the bubble interior is less sensitive to the jump in vacuum energy than the exterior. The reverse is true for (highly supressed) transitions from low to high curvature.  
This is a fundamentally important result. It tells us that in the low curvature vacua the vacuum energy contributions are most efficiently sequestered.  Intuitively this follows from the fact that an inertial observer is causally connected to a larger volume the {\it lower} the dS curvature -- once proper ratio of measures before and after the transition is set up by the junction conditions -- and as we have seen,  large volumes sequester most efficiently.
That vacuum energy in low curvature, near-Minkowski vacua are efficiently sequestered is hardly trivial: one could easily have imagined a scenario in which a near-Minkowski vacuum was only possible whenever the residual cosmological  constant was tuned against the jump in vacuum energy. However, thanks to the {\it dynamics} of sequestering such tuning is {\it not necessary}. If we take the residual cosmological constant to be small compared to the jump in vacuum energy, we are guaranteed to get a vacuum of low curvature!  This is in addition to the cancellation of the radiative corrections - sequestering here simply automatically protects the smallness of curvature in the vacua which start with small curvature to begin with, from any source of contamination.

This extends beyond dS to dS transitions, to the transitions describing including Minkowski $\rightarrow$ AdS. The latter are, of course,  complicated by the fact that gravitational collapse occurs inside an AdS  bubble. Generically, the vacuum with least absolute value of curvature is the one that is least sensitive to the jump in vacuum energy\footnote{There is one exception to this rule: tunnelling from a dS vacuum with large curvature, to an AdS vacuum with small (absolute) curvature. Then the ratio $\I$ is infinite meaning it is the large exterior dS curvature that is insensitive to the vacuum energy.  Indeed, there is a clear discontinuity between $dS \to $ Minkowski tunnelling and $dS \to AdS$ tunnelling, since gravitational collapse occurs inside the bubble of the latter.}.  In other words, generically,  no fine tuning against vacuum energy is required to achieve a low curvature vacuum in sequestering, even when tunnelling effects are taken into account. 

We conclude this section with a brief comment on  our choice of  regulator for the spacetime volume divergences: the expressions we derived for the volume ratios $\I_{X \to Y}$ above used a particular choice of volume regulator, as specified in the appendix. It was based on global coordinates and constant time slices. In particular, for dS space, we cut the volumes of at constant global time, and then take that cutoff to infinity. This choice guarantees that we cover the entire space and no more when the cut off is removed. The regulators in Minkowski and AdS space were chosen along the same lines.

\section{Summary}

Classical field theory is blind to the number of stable vacua with different energies. In quantum theory, dynamics of tunnelling permits transitions between different vacua, with well established semi-classical methods describing the transition via bubble nucleation. In a gravitational context such transitions result in a change of the spacetime curvature. The difference corresponds to the energy difference of the two vacua. 
Here we have considered such transitions, along the lines of Coleman and De Luccia \cite{ColemanDeLuccia}, only now in the context of the sequestering proposal \cite{KP1,KP2,KP3,KPSZ}.
In particular, we work with the manifestly local vacuum energy sequestering \cite{KPSZ}, where the radiative corrections to vacuum energy are 
automatically sequestered away from the gravitational field equations, rendering the vacuum curvature radiatively stable, in stark contrast to what happens in GR.  In the global sequestering it has been argued that the same occurs with vacuum energy transition from phase transitions \cite{KP1,KP2}.  Here we show this occurs in the local theory, where the dynamics of phase transitions occurs via bubble nucleation. Our main result is that, generically, the effects of the transition in vacuum energy are most efficiently sequestered in vacua with low absolute curvature, ie near Minkowski. This means that near-Minkowski vacua (eg dS with small curvature) are automatically safe from vacuum energy corrections induced by phase transitions, just as they are safe from quantum radiative corrections: {the locally small value of the vacuum curvature is  stable against matter loops {\it and} against the transition in vacuum energy}. 

To ensure absence of catastrophic instabilities in the theory, we must impose conditions (\ref{cond}) on the auxiliary functions controlling the couplings on the bare Planck scale and the bare cosmological constant to the topological sector which controls the sequester. Once we pick these couplings, tunnelling between vacua goes through in qualitatively the same way as GR: the allowed transitions generically lower the vacuum curvature. The one exception to this rule, the upwards transitions between dS vacua that are highly suppressed, as in GR. 
Curiously, the sequestering corrections render near-Minkowski vacua more stable than they would be in GR. Tunnelling from high dS curvature to Minkowski is enhanced relative to GR, while tunnelling from Minkowski to AdS is  suppressed relative to GR. There is no obvious reason to expect that this particular feature  will be generic to all adaptations of the sequestering proposal that exploit similar cancellation mechanisms. 

Sequestering effects become very significant for the determination of the local value of the vacuum curvature. Generically  it turns out that the effect of vacuum energy, and in particular the scale of the transition, is very efficiently sequestered in the vacuum region that dominates the spacetime volume. The opposite is true in the vacuum region with less spacetime volume. For the allowed configurations, this means that the vacuum energy contributions in near-Minkowski vacua are generically the most efficiently sequestered. We can understand this intuitively from the point of view of an inertial observer in dS space, whose static patch is larger the smaller the curvature (in the units of the Compton wavelength of a proton, say). So such observers do not have to fine tune the residual cosmological constant to cancel jumps in vacuum energy in order to protect their ``nearness" to Minkowski vacuum.  Note that this outcome is in stark contrast to GR where one always has to fine tune the bare cosmological constant against the transition scale in order to achieve a small vacuum curvature.

Our analysis of phase transitions in sequestering may open up a window to future tests of the proposal. For example, it was recently suggested that phase transitions in the interior of neutron stars could  affect their mass to size distribution, while cosmological phase transitions can affect the propagation of primordial gravitational waves \cite{terning}.  Phase transitions may also lead to detectable gravitational wave signals at ground and space based interferometers \cite{khoze}. As we have seen, 
phase transition contributions to vacuum energy are partially -- but very efficiently -- cancelled in sequestering, with the size of cancellation controlled by the relevant volume ratios. It would be interesting to consider the implications of these effects for concrete experimental searches.

\vskip.5cm

{\bf Acknowledgments}: 
N.K. is supported in part by the DOE Grant DE-SC0009999, AP was funded by a Royal Society URF, and DS by an STFC studentship.  


\appendix

\section{Maximally symmetric spacetimes}

Here we briefly review maximally symmetric spacetimes and their coordinate covers that were used in this paper. We begin with  dS space in $4$ dimensions which is best described as a hyperboloid embedded in $5$ dimensional Minkowski space. This makes the $SO(1,4)$ symmetry group of dS space manifest since the hyperboloid breaks the translational invariance of Minkowski space but leaves intact the Lorentz invariance.  dS space with a radius of curvature $1/q$ is described by the surface $W^{2} + R^{2} - T^{2} = 1/q^{2}$ embedded in 5D  Minkowski with the metric $ds_{5}^{2} = -dT^{2} + dW^{2} + dR^{2} + R^{2}d\Omega_{2}^{2}$ where $d\Omega_{2}^{2}$ describes a 2-sphere  ($d\Omega_{2}^{2} = d \chi^{2} + \sin^{2}\chi d\phi^{2}$). In global coordinates $(t,\theta, \chi, \phi)$ the metric on the hyperboloid is
\be
ds^{2} = -dt^{2} + \frac{\cosh^{2}qt}{q^{2}}(d\theta^{2} + \sin^{2}\theta d\Omega_{2}^{2}) \, ,
\ee
where $t \in (-\infty, \infty )$ and $\theta \in [0, \pi]$. The mapping from the embedding coordinates to the global coordinates on the hyperboloid is
\bea
W = \frac{1}{q} \cosh qt \cos \theta \, , \\
R = \frac{1}{q} \cosh qt \sin \theta \, , \\
T = \frac{1}{q} \sinh qt \, . 
\eea
Coleman's coordinates that only cover a patch of dS space are given by
\be \label{{colemancoord}}
ds^{2} = dr^{2} + \rho^{2}(r)(-d\tau^{2} + \cosh^2 \tau d\Omega_{2}^{2}) \, , 
\ee
where
\be
\rho(r) = \frac{\sin q(\epsilon r + r_{0})}{q} = \frac{\sin Q(r)}{q} \, ,
\ee
defining $Q(r)=q(\epsilon r + r_{0})$. Although these coordinates describe (a portion of) the same hyperboloid we define its mapping from the embedding space using different Minkowskian coordinates. This ensures that a point on the waist of the hyperboloid at  $\tau=0$ can be mapped to a point with arbitrary time in global coordinates. Since the hyperbolid is Lorentz invariant we can describe the same surface by performing a boost along the $W$ direction with rapidity $\alpha$. The global coordinates are then mapped to Coleman's by 
\beastar
\cos Q(r) = \cosh \alpha \cosh qt \cos \theta - \sinh \alpha \sinh qt \, , \\
\sin Q(r) \sinh \tau = \cosh \alpha \sinh qt - \sinh \alpha \cosh qt \cos \theta \, , \\
\sin Q(r) \cosh \tau = \cosh qt \sin \theta \, .
\eeastar
A similar description of AdS space exists,  however,  in 4 dimensions,  the symmetry group of the embedding manifold is $SO(2,3)$ rather than $SO(1,4)$. Therefore AdS space is described by the surface $W^{2} + T^{2} - R^{2} = 1/|q|^{2}$ embedded in $ds_{5}^{2} = -dT^{2} - dW^{2} + dR^{2} + R^{2}d\Omega_{2}^{2}$. In global coordinates $(t,u,\chi,\phi)$ the metric on the hyperboloid is
\be
ds^{2} = - \frac{\cosh^{2} |q| u}{|q|^{2}} dt^{2} + du^{2} + \frac{\sinh^{2}|q| u}{|q|^{2}} d\Omega_{2}^{2} \, ,
\ee
where $t\in (-\infty,\infty)$, $u \in [0, \infty)$ and the mapping from the embedding coordinates to the global ones is
\bea
W = \frac{1}{|q|} \cosh |q| u \cos t \, ,  \\
T = \frac{1}{|q|} \cosh |q| u \sin t \, , \\
R = \frac{1}{|q|} \sinh |q| u \, .
\eea
Using Coleman's coordinates  the metric is equivalent to equation (\ref{{colemancoord}}) with
\be
\rho(r) = \frac{\sinh |q| (\epsilon r + r_{0})}{|q|} = \frac{\sinh Q(r)}{|q|}
\ee
where now $Q(r)=|q|(\epsilon r + r_{0})$. Ensuring that $\tau=0$ maps to an arbitrary global time, we make use of the $SO(2,3)$ symmetry in the embedding manifold and perform a rotation in the $T-W$ plane before mapping to Coleman's coordinates. The relevant mapping is now
\bea
 \label{adsmap}
\cosh Q(r) = \cosh |q| u \cos(t -t_0) \, , \\
\sinh Q(r) \sinh \tau = \cosh |q|u \sin(t -t_0) \, , \\
\sinh Q(r) \cosh \tau = \sinh |q|u \, ,
\eea
where $t_0$ is the angle of rotation. 

The remaining maximally symmetric spacetime is Minkowski space with a vanishing cosmological constant. In terms of Coleman's coordinates this corresponds to the same metric of equation (\ref{{colemancoord}}) but with $\rho(r) = \epsilon r + r_{0}$, and in global coordinates has the usual flat metric $ds^{2} = -dt^{2} + du^{2} + u^{2}d\Omega^2_{2}$. The mapping from Coleman's coordinates to global coordinates is given by
\bea
u = \rho(r) \cosh \tau \, ,\\
t = \rho(r) \sinh \tau + t_{0} \, .
\eea
Again, in order that $\tau=0$ maps to an arbitrary global time we have used the translational invariance of Minkowski space to shift the temporal coordinate before defining the mapping.

\section{Location of AdS singularity}

As shown in \cite{AbbottColeman}, an AdS interior suffers from a curvature singularity due to the collapse of the bubble in the future. So we need to cut of the spacetime at this surface. Here we calculate where this surface is in global coordinates. A coordinate singularity occurs at $Q(r) = 0$ in Coleman's coordinates so at this point we change to cosmological coordinates with
\be
ds^{2} = -d \eta^{2} + \bar{\rho}^{2}(\eta)(d \lambda^{2} + \sinh^{2} \lambda d \Omega_{2}^{2}) \, , 
\ee
where $\bar{\rho}(\eta) = {\sin |q| \eta}/{|q|}$. Again we can describe this spacetime as an embedding in a 5D spacetime with symmetry $SO(2,3)$. The embedding is
\bea
\bar{W} = \frac{\cos |q| \eta}{|q|} \, , \\
\bar{T} = \frac{\sin |q| \eta}{|q|} \cosh \lambda \, , \\
\bar{R} = \frac{\sin |q| \eta}{|q|} \sinh \lambda \, .
\eea
From \cite{AbbottColeman}  a curvature singularity forms at $\eta = \pi/|q|$. We use the same embedding coordinates when mapping to the cosmological ones as we do when mapping to Coleman's such that
\bea
\cosh Q(r) = \cos |q| \eta \, , \\ \label{colemantocosmo}
\sinh Q(r) \sinh \tau = \sin |q| \eta \cosh \lambda \, , \\
\sinh Q(r) \cosh \tau = \sin |q| \eta \sinh \lambda \, .
\eea
From equation (\ref{colemantocosmo}) the singularity is at $\cosh Q(r) = -1$ and so in global coordinates the curvature singularity corresponds to the surface $\cosh |q|u \cos(t-t_0) = -1$ by equation (\ref{adsmap}).

\section{Spacetime volumes}

The volume of dS space in global coordinates is
\be
\frac{V^{dS}_{\text{total}}}{\Omega_{2}} = \int^{\infty}_{-\infty} dt \frac{\cosh^{3} qt}{q^{3}} \int_{0}^{\pi} d\theta \sin^{2} \theta  \, .
\ee
The volume of AdS space in global coordinates is
\be
\frac{V^{AdS}_{\text{total}}}{\Omega_{2}} = \int^{\infty}_{-\infty} dt  \int^{\infty}_{0} du \frac{\cosh |q|u \sinh^{2}|q|u}{|q|^{3}} \, .
\ee
The volume of Minkowski space in global coordinates is
\be
\frac{V^{M}_{total}}{\Omega_{2}} = \int_{-\infty}^{\infty} dt \int_{0}^{\infty} du u^{2} \, .
\ee
Each volume is divergent and so they need to be regulated on the relevant surface. We only care about the ratio of the volumes so the divergent behavior in each case is sufficient to determine the behavior of the ratios.

\begin{figure}[thb] 
  \centering
  \includegraphics[width=3in]{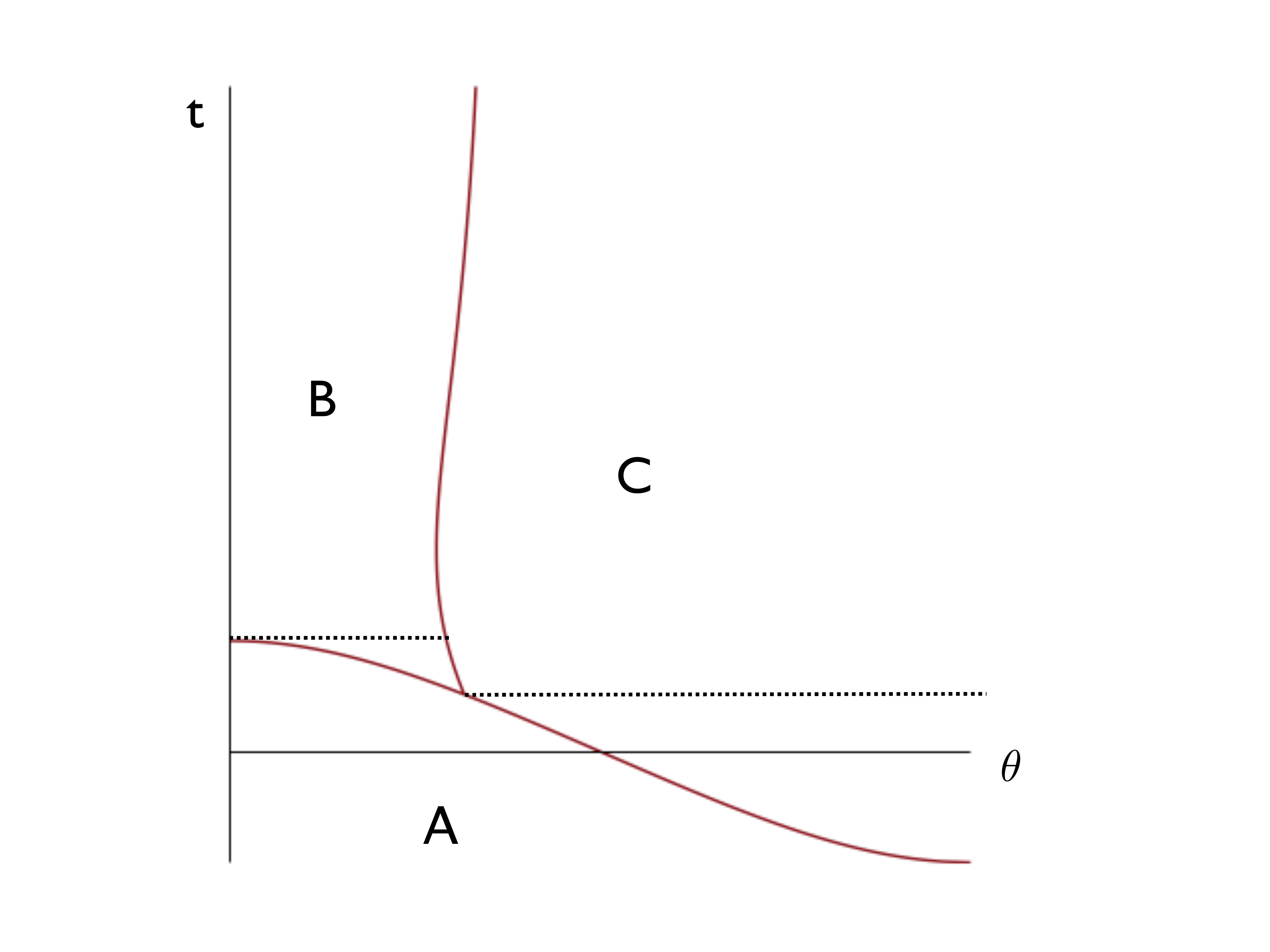} 
  \caption{Slicing of spacetime volumes in global coordinates for tunnelling involving dS vacua. Region A described the era before nucleation of a bubble, while regions B and C  can be either the interior  or  exterior of a bubble depending on the configuration. Regions B and C are each split into two parts by a dotted black line  to make the integrals simpler to handle. }
    \label{tundsds}
\end{figure}

There are generic regions of spacetime that appear in each of our bubble geometries. In dS, there are three volumes that may be of interest, as shown in Figure \ref{tundsds}.
Let us compute these volumes: for $X=A, B ,C$, we have 
\be
\frac{V^{dS}_X}{\Omega_{2}} = \int_{X} dt d\theta  \frac{\cosh^{3} q t \sin^{2} \theta}{q^{3}}\, .
\ee
The nucleation surface representing the boundary of region A corresponds to  time $\tau = 0$ in Coleman's coordinates and the surface $\tanh qt = \tanh \alpha \cos \theta$ in global coordinates, ie
\be
t_{tun}(\theta)=\frac1q \tanh^{-1} \left( \tanh \alpha \cos \theta \right) \, .
\ee
The wall separating regions B and C is at $r=0$ and corresponds to the surface $\cos q r_{0} = \cosh \alpha \cosh q  t \cos \theta - \sinh \alpha \sinh q t$ in global coordinates, ie
 \be
 \theta_{wall}(t)=\cos^{-1} \left( \frac{\cos q r_{0} +  \sinh \alpha \sinh q t}{\cosh \alpha \cosh q  t } \right) \, .
\ee
The volume of region A diverges as $t  \to  -\infty$ and we must regulate it, cutting the spacetime off at $t=-t_{reg}$. This region of spacetime is defined up until bubble nucleation and so the upper limit of the $t$ integral is given by the nucleation surface $t=t_{tun}(\theta)$, which  is finite for finite $\alpha$. The volume is 
\be
\frac{V^{dS}_A}{\Omega_{2}} = \int_{0}^{\pi} d\theta \sin^{2} \theta \int_{-t_{reg}}^{t_{tun}(\theta)} dt \frac{\cosh^{3} q t}{q ^{3}} \, .
\ee
Performing the $t$ integral we have
\be
\frac{V^{dS}_A}{\Omega_{2}} = \int_{0}^{\pi} d\theta \frac{\sin^{2} \theta}{3 q^{4}}[\sinh qt (\cosh^{2} qt + 2)]^{t_{tun}(\theta)}_{-t_{reg}} \, .
\ee
Since $t_{tun}$ is finite, the integral is dominated by the regulated surface and can therefore be approximated by
\be
\frac{V^{dS}_A}{\Omega_{2}} \sim \frac{\pi}{48 q ^{4}} e^{3 q t_{reg}} \, .
\ee

We now consider the volumes after bubble nucleation. For region B we have
\be
\frac{V^{dS}_B}{\Omega_{2}} =  \int^{\theta_{wall}(t)}_{0} d \theta \sin^{2} \theta \int^{t_{reg}}_{t_{\dagger}} dt \frac{\cosh^{3} qt}{q^{3}} \, ,
\ee
where we have ignored the contribution from below the dotted black line (at $t=t_\dagger$)  since it is finite even as we remove the regulator. Note that the late time regulator at $t=t_{reg}$ is assumed to be equal in size to the early time regulator $t=-t_{reg}$.  This choice guarantees that, in the absence of a wall, the volume in $t \geq0$ is exactly equal to the volume in $t \leq 0$.  Doing the integrals,
\be
\frac{V^{dS}_B}{\Omega_{2}} \sim \left(\cos^{-1}(\tanh \alpha) - \frac{\tanh \alpha}{\cosh \alpha}\right) \frac{1}{48 q^{4}} e^{3 qt_{reg}} \, .
\ee
For region C, 
\be
\frac{V^{dS}_C}{\Omega_{2}} =  \int^{\pi}_{\theta_{wall}(t)} d\theta \sin^{2} \theta  \int_{t_{\star}}^{t_{reg}} dt \frac{\cosh^{3} qt}{q^{3}} \, .
\ee
Again, we dropped the contribution from below the dotted black line (at $t=t_\star$)  since it is finite even as we remove the regulator.  Performing the integrals yields
\be
\frac{V^{dS}_C}{\Omega_{2}} \sim \left(\pi - \cos^{-1}(\tanh \alpha) + \frac{\tanh \alpha}{\cosh \alpha}\right) \frac{1}{48 q^{4}} e^{3 qt_{reg}} \, .
\ee
Note that $V_A^{dS} \sim V_B^{dS}+V_C^{dS} \sim {V_{total}^{dS}}/{2}$

\begin{figure}[thb] 
  \centering
  \includegraphics[width=3in]{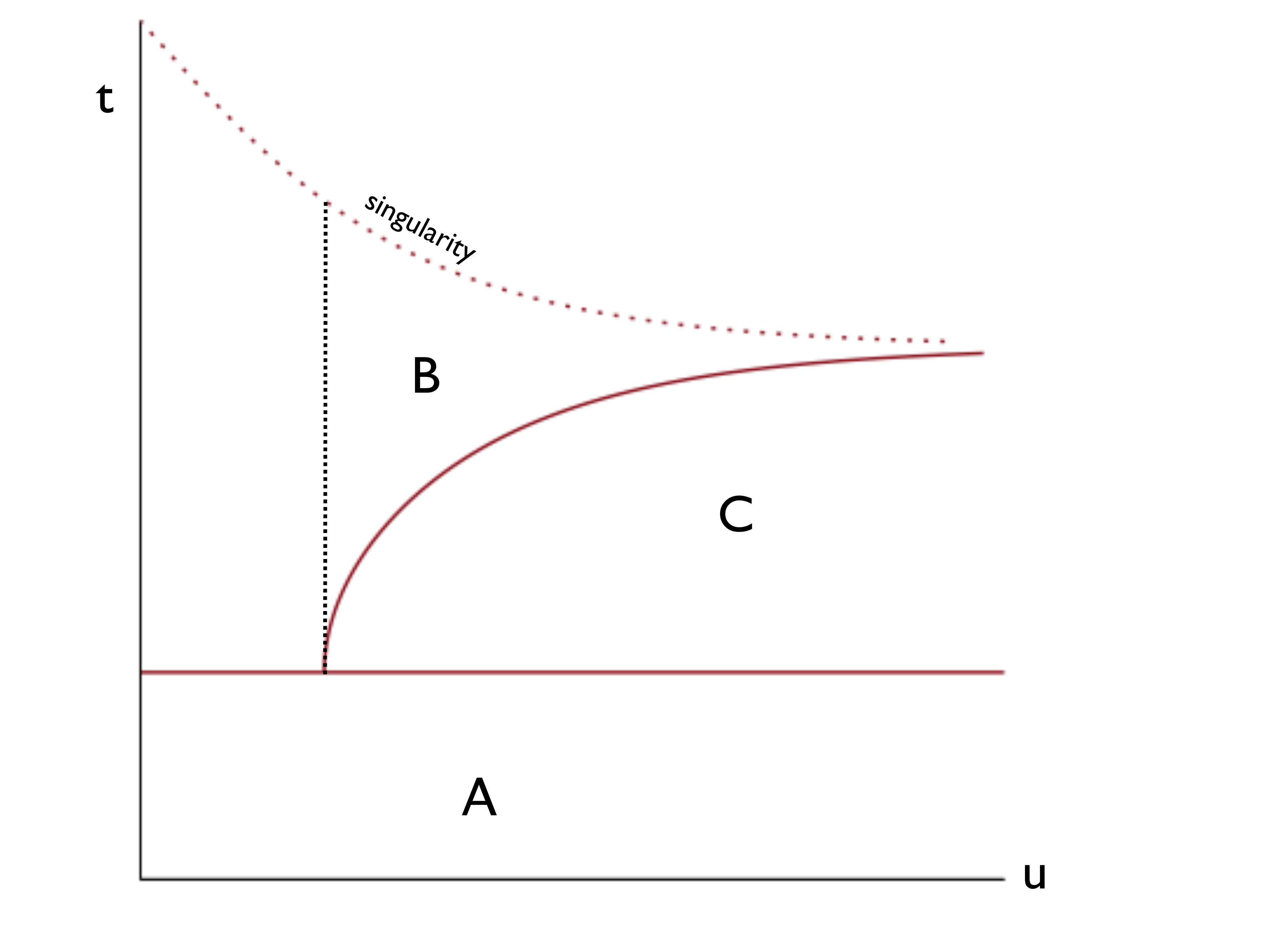} 
  \caption{Slicing of spacetime volumes in global coordinates for tunnelling involving  AdS vacua. Region A corresponds to the time before nucleation of a bubble, region B to a bubble interior and region C  to a  bubble exterior for the allowed configurations. Region B  is split into two parts by a dotted black line  to make the integrals  simpler to handle, and is also cut off by a singularity. }
    \label{tunadsads}
\end{figure}

Now consider the relevant  sections of AdS space, as shown in Figure \ref{tunadsads}.
Again, we have three volumes of interest, for $X=A,  B, C$
\be
\frac{V^{AdS}_X}{\Omega_{2}} = \int_{X} dt du \frac{\cosh |q| u \sinh^{2} |q|u }{|q|^{3}}\, .
\ee
The nucleation surface corresponding to the boundary of region A is $\tau = 0$, which in global coordinates this translates to
$
\cosh |q| u \sin (t -t_0) = 0
$.
Taking the principle root for definiteness,  we have  $t =t_0$ . The wall separating regions B and C is at $r = 0$ which in global coordinates corresponds to
$
\cosh |q| r_{0} = \cosh |q|u \cos (t-t_0)
$, ie 
\be
t_{wall}(u) =t_0+\cos^{-1} \left( \frac{\cosh |q| r_{0}}{ \cosh |q|u} \right) \in (t_0, t_0+\frac{\pi}{2} ) \, .
\ee
For tunnelling into AdS there is an added complication due to the singularity in the AdS interior in region B. In the previous section we showed that this lay at
\be
t_{sing}(u)=t_0^-+\cos^{-1} \left(\frac{-1}{\cosh |q|_-u} \right) \in (t_0^-+\frac{\pi}{2} , t_0^-+\pi) \, .
\ee
Consider first the volume of the region A, before bubble nucleation. This is given by
\be
\frac{V_A^{AdS}} {\Omega_2}= \int^{t_0}_{-\infty} dt \int_{0}^{u_{reg}} du \frac{\cosh |q|u \sinh^{2} |q| u}{|q|^{3}} \, .
\ee
This has two divergent directions. For now we regulate in $u$ only, cutting off the integral at $u=u_{reg}$, such that 
\be
\frac{V_A^{AdS}} {\Omega_2}\sim   \frac{1}{24|q|^{4}} e^{3 |q| u_{reg}}   \int^{t_0}_{-\infty} dt \, .
\ee

Now consider region B, which corresponds to the interior of an AdS bubble. Cutting off the AdS space at the singularity surface we find that the volume is given by 
\be
\frac{V_B^{AdS} }{\Omega_{2}} =\int_{u_\star}^{u_{reg}} du \frac{\cosh |q| u \sinh^{2} |q|u}{|q| ^{3}} \int_{t_{wall}(u)}^{t_{sing}(u)} dt \, ,
\ee
where we have neglected the finite part to the left of dotted black line (at $u=u_\star$).  We regulate the divergence in the integral by cutting it off at $u=u_{reg}$. This yields a leading order contribution
\be
\frac{V_B^{AdS}}{\Omega_{2}}  \sim \frac{(1+\cosh |q| r_0)}{8 |q|^4} e^{2|q| u_{reg}} \, .
\ee
For region C, corresponding to the an AdS exterior, we have
\be
\frac{V_C^{AdS}}{\Omega_{2}} = \int^{u_{reg}}_{u_{wall}(t_0)} du  \frac{\cosh |q|u \sinh^{2} |q|u}{|q|^{3}}  \int^{t_{wall}(u)}_{t_0} dt \, ,
\ee
where again   we have regulated the $u$ integral by cutting it off at $u=u_{reg}$. We approximate this volume as
\be
\frac{V_C^{AdS}}{\Omega_{2}} \sim \frac{\pi}{ 48 |q|^{4}} e^{3|q|u_{reg}}  \, .
\ee

\begin{figure}[thb] 
  \centering
  \includegraphics[width=3in]{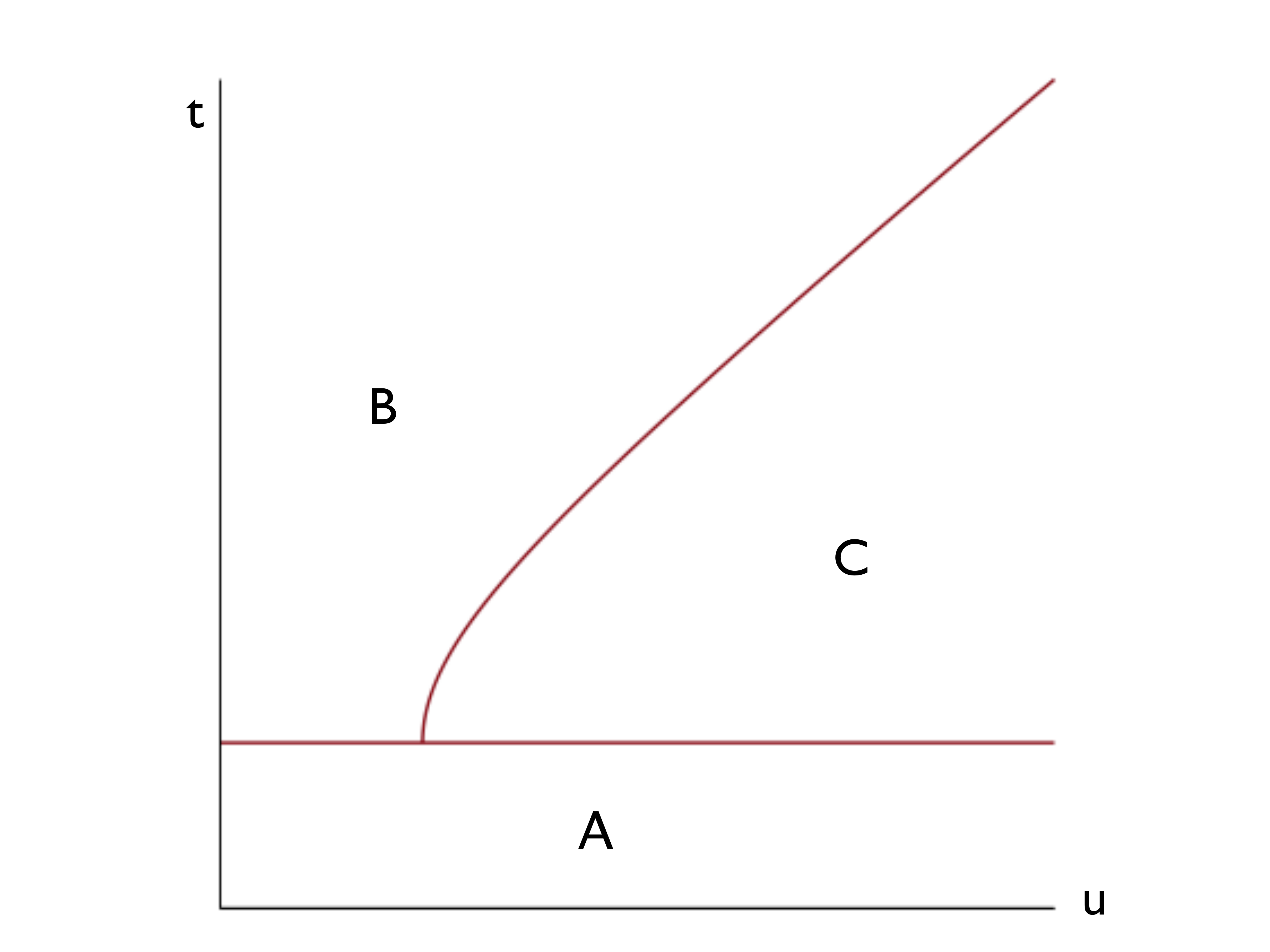} 
  \caption{Dissection of spacetime volumes in global coordinates for tunnelling involving  Minkowski vacua. Region A corresponds to the time before nucleation of a bubble, region B to a bubble interior and region C  to a  bubble exterior for the allowed configurations. }
    \label{tunMM}
\end{figure}

Finally we turn to the relevant sections of Minkowski space, as shown in Figure \ref{tunMM}.
Again, we have three volumes to calculate, for $X=A, B, C$
\be
\frac{V^{M}_X}{\Omega_{2}} = \int_{X} dt du u^2 \, .
\ee
Tunnelling can occur at the boundary of region A, which is given by global time $t=t_0$. The wall separating $B$ and $C$ is given by 
$u_{wall}(t)=\sqrt{r_0^2+(t-t_0)^2}$, or equivalently
$t_{wall}(u)=t_0+\sqrt{u^2 -r_0^2}$. 
The initial volume, of region A, is
\be
\frac{V_A^{M}} {\Omega_2}= \int^{t_0}_{-\infty} dt \int_{0}^{u_{reg}} du u^2 \, .
\ee
This has two divergent directions, but for now we only regulate the $u$ integral, at $u=u_{reg}$, yielding
\be
\frac{V_A^{M}} {\Omega_2} \sim  \frac{u_{reg}^3}{3}  \int^{t_0}_{-\infty} dt \, .
\ee
For region $B$ we must regulate the time integral, cutting it off at $t=t_{reg}$, such that
\be
\frac{V_B^{M}} {\Omega_2}= \int_{t_0}^{t_{reg}} dt \int_{0}^{u_{wall}(t) } du u^2 \sim \frac{t_{reg}^4}{12} \, .
\ee
Finally, for region C, we regulate the $u$ integral, just as we did for region A, yielding
\be
\frac{V_C^{M}} {\Omega_2}=  \int^{u_{reg}}_{u_{wall}(t_0) } du u^2\int_{t_0}^{t_{wall}(u)} dt  \sim \frac{u_{reg}^4}{4} \, .
\ee

\section{Volume ratios}

We shall now compute the ratio $\I =\frac{\V_{b}^{+} + \V_{a}^{+}}{\V_{a}^{-}}$ for each of the allowed configurations. We start with tunnelling between dS vacua, for which there are four configurations, ultimately corresponding to the four possible arrangements of the geometries after bubble nucleation (ie BB, BC, CB, CC). The summary of results is
\bea
\V_b^+ &\sim&   \frac{\pi}{48 q_+ ^{4}} e^{3 q_+ t_{reg}^+} \, , \\ 
\V_a^+&\sim& f_+(\alpha_+)\frac{\pi}{48 q_+ ^{4}} e^{3 q_+ t_{reg}^+} \, , \\
\V_a^-&\sim& f_-(\alpha_-)\frac{\pi}{48 q_- ^{4}} e^{3 q_- t_{reg}^-} \, , 
\eea
where the form of the coefficients $f_\pm$ depends on the chosen orientation (ie BB, BC, CB or CC). To relate the regulators inside and out the bubble we match the geometries at their point of intersection with the wall. In other words, we match the radius of the 2-sphere at this intersection point, $\frac1q \cosh q t_{reg} \sin \theta_{wall}(t_{reg})$, yielding
\be
\left(\frac{e^{q t_{reg}}}{q \cosh \alpha}\right)_{+} = \left(\frac{e^{q t_{reg}}}{q \cosh \alpha}\right)_{-} \, ,
\ee
at large values of the $t_{reg}$. This equation is just the Israel junction condition on the surface of the bubble in the angular directions.

This now allows us to compare the volume terms inside and outside of the bubble, ultimately yielding a ratio of the form
\be
\I_{dS \to dS} \sim \frac{\cosh^{3} \alpha_{+}}{\cosh^{3} \alpha_{-}} \left(\frac{1 + f_+(\alpha_+)}{f_-(\alpha_-)}\right) \frac{q_-}{q_+} \, .
\ee
We neglect possible  zeros or singular points in the coefficient of $\frac{q_-}{q_+}$ since these will only occur for very precise nucleation times, and are not generic. 

Now consider tunnelling between AdS  vacua. From the previous section, we now have
\bea
\V_b^+ &\sim&  \frac{1}{24|q|_+^{4}} e^{3 |q|_+ u^+_{reg}}   \int^{t^+_0}_{-\infty} dt  \, , \\ 
\V_a^+&\sim& \frac{\pi}{ 48 |q|_+^{4}} e^{3|q|_+u^+_{reg}} \, , \\
\V_a^-&\sim&   \frac{(1+\cosh |q|_- r_0^-)}{8 |q|_-^4} e^{2|q|_- u^-_{reg}} \, .
\eea
Matching the radius of the 2-sphere at the intersection of the regulators with the wall now gives
\be
\left(\frac{e^{|q| u_{reg}}}{|q|}\right)_{+} = \left(\frac{e^{|q| u_{reg}}}{|q| }\right)_{-} \, ,
\ee
for large $u_{reg}$. Using this to compute the corresponding  ratio gives
\be
\I_{{AdS} \to {AdS}} \to \infty \, .
\ee

For tunnelling from dS into AdS, we have
\bea
\V_b^+ &\sim&   \frac{\pi}{48 q_+ ^{4}} e^{3 q_+ t_{reg}^+} \, , \\ 
\V_a^+&\sim& f_+(\alpha_+)\frac{\pi}{48 q_+ ^{4}} e^{3 q_+ t_{reg}^+} \, , \\
\V_a^-&\sim&   \frac{(1+\cosh |q|_- r_0^-)}{8 |q|_-^4} e^{2|q|_- u^-_{reg}} \, .
\eea
Matching the radius of the 2-sphere at the intersection of the regulators with the wall gives
\be
\left(\frac{e^{q t_{reg}}}{q \cosh \alpha}\right)_{+} = \left(\frac{e^{|q| u_{reg}}}{q }\right)_{-} \, ,
\ee
for large $t_{reg}$ and $u_{reg}$. This implies that the ratio is
\be
\I_{{dS} \to {AdS}} \to \infty \, ,
\ee
where we once again neglect any non-generic behavior occurring at special isolated choices of $\alpha_+$.

We now turn to transitions to Minkowski vacua. For tunnelling from dS into Minkowski we have
\bea
\V_b^+ &\sim&   \frac{\pi}{48 q_+ ^{4}} e^{3 q_+ t_{reg}^+}\, , \\
\V_a^+&\sim& f_+(\alpha_+)\frac{\pi}{48 q_+ ^{4}} e^{3 q_+ t_{reg}^+}\, , \\
\V_a^-&\sim&  \frac{(t^-_{reg})^4}{12} \, .
\eea
Matching the regulators now gives
\be
\left(\frac{e^{q t_{reg}}}{2q \cosh \alpha}\right)_{+} = t_{reg}^- \, ,
\ee
which gives
\be
\I_{{dS} \to {M}} = 0 \, .
\ee
As before we ignore special cases that might occur at special isolated choices of $\alpha_+$.

Finally we consider tunnelling from Minkowski into AdS, for which 
\bea
\V_b^+ &\sim&  \frac{ (u^+_{reg})^3} {3}  \int^{t^+_0}_{-\infty} dt  \, , \\ 
\V_a^+&\sim&  \frac{ (u^+_{reg})^4} {4} \, ,  \\
\V_a^-&\sim&   \frac{(1+\cosh |q|_- r_0^-)}{8 |q|_-^4} e^{2|q|_- u^-_{reg}} \, .
\eea
Matching the regulators we find
\be
u_{reg}^+= \left(\frac{e^{|q| u_{reg}}}{2|q| }\right)_{-} \, ,
\ee
and so
\be
\I_{{M} \to {AdS}} \to \infty \, .
\ee

\end{document}